\documentclass[%
prx,
 amsmath,amssymb,
 aps,
twocolumn,
longbibliography,
showkeys
]{revtex4-1}

\usepackage{graphicx}
\usepackage{epstopdf}
\usepackage{dcolumn}
\usepackage{bm}
\usepackage{color}
\usepackage[mathlines]{lineno}
\usepackage[breaklinks=true,colorlinks=true,linkcolor=blue,urlcolor=blue,citecolor=blue]{hyperref}
\usepackage{ragged2e}
\usepackage{enumitem}
\usepackage{bibunits}
\usepackage{amsthm}
\usepackage{diagbox}

\renewcommand\thefigure{\arabic{figure}}

\begin{document}


\title[]{Cooperative bots exhibit nuanced effects on cooperation across strategic frameworks}

\author{Zehua Si$^1$}
\author{Zhixue He$^{2,1}$}
\author{Chen Shen$^3$}
\email{steven\_shen91@hotmail.com}
\author{Jun Tanimoto$^{1,3}$}
\email{tanimoto@cm.kyushu-u.ac.jp}

\affiliation{
\vspace{2mm}
\mbox{1. Interdisciplinary Graduate School of Engineering Sciences, Kyushu University, Fukuoka, 816-8580, Japan}
\mbox{2. School of Statistics and Mathematics, Yunnan University of Finance and Economics, Kunming, 650221, China}
\mbox{3. Faculty of Engineering Sciences, Kyushu University, Kasuga-koen, Kasuga-shi, Fukuoka 816-8580, Japan}
}

\date{\today}

\begin{abstract}
The positive impact of cooperative bots on cooperation within evolutionary game theory is well documented; however, existing studies have predominantly used discrete strategic frameworks, focusing on deterministic actions with a fixed probability of one. This paper extends the investigation to continuous and mixed strategic approaches. Continuous strategies employ intermediate probabilities to convey varying degrees of cooperation and focus on expected payoffs. In contrast, mixed strategies calculate immediate payoffs from actions chosen at a given moment within these probabilities. Using the prisoner's dilemma game, this study examines the effects of cooperative bots on human cooperation within hybrid populations of human players and simple bots, across both well-mixed and structured populations. Our findings reveal that cooperative bots significantly enhance cooperation in both population types across these strategic approaches under weak imitation scenarios, where players are less concerned with material gains. However, under strong imitation scenarios, while cooperative bots do not alter the defective equilibrium in well-mixed populations, they have varied impacts in structured populations across these strategic approaches. Specifically, they disrupt cooperation under discrete and continuous strategies but facilitate it under mixed strategies. These results highlight the nuanced effects of cooperative bots within different strategic frameworks and underscore the need for careful deployment, as their effectiveness is highly sensitive to how humans update their actions and their chosen strategic approach.
\end{abstract}


\maketitle

\section{\label{sec:level1}Introduction}
Cooperation, which incurs a cost to benefit others, presents a paradox, as helping others should reduce one's own benefits and lead to extinction according to the principle of ``survival of the fittest''~\cite{darwin1964origin,axelrod1981evolution,rand2013human}. Yet, this prediction contrasts with the ubiquitous cooperation observed in both nature and human societies~\cite{dugatkin1997cooperation,fowler2010cooperative,perc2017statistical}. Traditionally, research on this conundrum has focused on biological altruism, developing theories of reciprocity~\cite{rand2013human,nowak2006five} and prosocial preferences~\cite{fehr2002strong,fehr2018normative,wang2020communicating} to explain observed cooperative behaviors. However, with the advancement of artificial intelligence (AI), humans are increasingly interacting with social bots in various realms~\cite{varol2017online,grimme2017social,shirado2020network}. For example, in customer service, chatbots on retail websites provide personalized shopping advice, facilitating effective problem-solving cooperation. In healthcare, virtual assistants offer support for scheduling appointments and providing medical information, enhancing cooperative patient care. On social media platforms, bots engage with users to simulate social interactions and influence opinions, impacting social dynamics and group behaviors. As AI becomes more integrated into human interactions, it progressively reshapes our understanding of how cooperation evolves and is sustained in a hybrid population of humans and AI.

In the framework of evolutionary game theory~\cite{smith1982evolution,weibull1997evolutionary,sigmund1999evolutionary}, current research on cooperation within human-AI systems unfolds from two primary perspectives. The first focuses on developing algorithms to achieve human-level cooperation~\cite{crandall2018cooperating, oudah2018ai, zhang2023rethinking}, exploring the phenomenon of human bias or machine penalty---the reluctance to cooperate with machines compared to playing with humans~\cite{karpus2021algorithm, ishowo2019behavioural, melo2016people}, and developing methods to mitigate or overcome this machine penalty~\cite{andras2018trusting,bonnefon2024moral}. The other perspective explores the role of AI as a scaffold for human cooperation, acting in various capacities such as planners structuring network interactions~\cite{shirado2020network,mckee2023scaffolding}, independent decision-makers affecting population composition~\cite{terrucha2024art,guo2023facilitating,sharma2023small,han2021or}, and proxies making decisions on behalf of humans~\cite{terveen1995overview,de2019cooperation,fernandez2022delegation}. For a comprehensive understanding, ref. \cite{guo2024multi} provides a thorough review. 

Research involving AI as independent decision-makers usually occurs within the context of one-shot and anonymous games. In such scenarios, players do not interact with the same opponents more than once and lack information about their opponents in each game round. It is straightforward to design AI algorithms with fixed actions or to leverage self-regarding reinforcement learning algorithms in such settings~\cite{santos2019evolution,wang2022modelling,sharma2023small,shen2024prosocial,shi2024enhancing}. These studies consistently show that cooperative bots are capable of assisting humans in solving complex problems that are difficult to solve alone, including cooperation challenges~\cite{shirado2020network, sharma2023small}, collective risk dilemmas~\cite{fernandez2022delegation, terrucha2024art}, and the punishment puzzle~\cite{shen2024prosocial}. However, these human-agent studies often rely on a discrete strategic approach, which only allows for deterministic actions with a fixed probability of one. While this approach is useful for modeling, it tends to oversimplify and idealize the complexities inherent in human decision-making processes: On one hand, for instance in investment behaviors, the changes in human strategies (investment amounts) are a continuous process~\cite{killingback2002continuous}; on the other hand, although humans' inherent cooperative tendencies variate continuously, in some scenarios, they have to choose a specific action between cooperation and defection based on their cooperative inclinations during the game. For example, in an election campaign with two candidates, each person's preference for the two candidates varies, but ultimately they must choose one of the two.

In this paper, we move beyond the traditional discrete strategic framework, commonly applied to studies of AI with fixed behaviors in cooperation, and expand our exploration into continuous and mixed strategic approaches. Unlike the discrete strategy, which relies on binary choices, the continuous strategic approach utilizes intermediate probabilities to convey varying degrees of cooperation and focuses on calculating expected payoffs~\cite{mar1994chaos,frean1996evolution,roberts1998development,killingback2002continuous,wahl1999continuous,wahl1999continuous1}. In contrast, the mixed strategic approach---while similar to the continuous strategy in employing probabilistic decisions---differs by calculating the immediate payoff of the action chosen at that moment, instead of an expected payoff~\cite{kokubo2015spatial,tanimoto2015correlated}. In exploring these strategies within hybrid populations of human players and simple bots, we use the prisoner's dilemma game~\cite{rapoport1965prisoner,tanimoto2019evolutionary}, where players can either cooperate, benefit others, or defect for self-interest. Within the discrete strategic framework, players opt for unconditional cooperation or defection, each with a fixed probability of one. Meanwhile, in the continuous and mixed approaches, players choose their cooperation level based on certain probabilities that guide their actions. We assume human players update their strategies through social learning by imitating the most successful strategy—i.e., the one yielding the highest payoff. The simple bots, however, are pre-designed to consistently exhibit high cooperative tendencies. Additionally, we explore two representative population structures: a well-mixed population, where players interact with others at equal probability, and a structured population represented as two-dimensional regular lattice network, where interactions are limited to direct neighbors.

Our results demonstrate that introducing cooperative bots under these three strategic approaches can significantly enhance cooperation among human players in both well-mixed and structured populations under a weak imitation scenario, where players are less concerned with their material gains. Conversely, under a strong imitation scenario, where players closely emulate the most successful strategies, the introduction of cooperative bots does not alter the prevailing defective equilibrium in well-mixed populations. This outcome holds across all three strategic approaches. However, the influence on cooperation in structured populations shows distinct patterns: cooperative bots disrupt cooperation among human players under both discrete and continuous strategic approaches, yet they still facilitate cooperation under the mixed strategic approach. These findings highlight nuanced differences in the impact of cooperative bots on human cooperation, deepening our understanding of how AI with fixed behaviors influences cooperation among human players within the framework of evolutionary game theory.

\section{Model}
Our model approach is structured around four key components: A. discrete, continuous, and mixed strategies setups; B. population settings; C. strategy update; and D. simulation settings. Concise explanations of each component will follow in sequence.

\subsection{\label{sec:level2} Discrete, continuous, and mixed strategies setups}
We employ the prisoner's dilemma (PD) game as the paradigm and consider three distinct strategies in our study:

In the discrete strategic approach, two players each make a choice between pure cooperation (C) or pure defection (D), with the probability of 1. 
If both players choose cooperation, each of them receives a reward $R$, whereas mutual defection results in a punishment $P$. If one player cooperates while the other defects, the cooperator earns a sucker's payoff $S$, and the defector gains a temptation payoff $T$. Based on the concept of dilemma strength outlined in Refs.~\cite{wang2015universal,tanimoto2021sociophysics}, we defined the chicken-type dilemma as $D_g = T-R$ and the stag hunt-type as $D_r = P-S$. For the sake of simplicity and without loss of generality, we set $P = 0$ and $R = 1$, which allows us to represent the payoff matrix as~\ref{eq1}:

\begin{equation}
{
\left[ \begin{array}{cc}
R & S \\
T & P 
\end{array} 
\right ]}={
\left[ \begin{array}{cc}
1 & -D_r \\
1+D_g & 0 
\end{array} 
\right ]}.
\label{eq1}
\end{equation}

This representation adheres to the restrictions $T > R > P > S$ and $2R > T+S$. If we assume $D_g=D_r=r\in[0,1]$, the dilemma strength of the PD game can be succinctly represented by the single parameter, $r$.

In contrast to the discrete strategic approach where the options for players are limited to either $C$ or $D$, each with a fixed probability of 1, a continuous strategy framework permits players to choose $C$ with any probability $s$ within the interval $[0,1]$, and correspondingly $D$ with the probability $1-s$. This allows for a range of choices that span from purely cooperative to purely defecting, rather than just a binary choice. Under such an approach, the payoff for player $i$, as outlined in Eq.~\ref{eq2}, is determined by computing the expected payoff using the parameters defined in the payoff matrix from~\ref{eq1}.

\begin{equation}
\begin{split}
\pi_{ij}=-D_r\cdot s_i+(1+D_g)\cdot s_j
&+(-D_g+D_r)\cdot s_i\cdot s_j\\
&=-r\cdot s_i+(1+r)\cdot s_j
\end{split}
\label{eq2}
\end{equation}

In the mixed strategic approach, akin to the continuous strategic approach, players select a strategy $s$ that ranges from 0 to 1. However, unlike the continuous strategic approach where $s$ simultaneously denotes the player's strategy and action, the mixed strategic approach necessitates that players explicitly decide on an action between $C$ and $D$ based on their chosen strategy. Consequently, the action spectrum available to players is binary, mirroring the discrete strategic approach rather than continuous ones. The computation of their payoffs adheres to the same methodology applied in discrete strategies (see Eq.~\ref{eq1}).

\subsection{\label{sec:level2}Population structures}
We consider two distinct player types: bot players (BP) and ordinary players (OP). The distinction between them lies in the fact that the latter update their strategies by imitating other players, while the former do not. In other words, bots consistently maintain their initial strategy unchanged throughout multiple rounds of the game. We assume that the proportion of bots in the hybrid population is denoted by $\rho$. Since we aim to investigate whether a small number of bots can promote cooperation among ordinary players, we assume that the proportion of bots in the hybrid population does not exceed 0.5. Furthermore, we assume that all bots share the same strategic value $s_{BP}=\theta\in[0,1]$ under continuous and mixed strategic approaches. In the case of the discrete strategic approach, $\theta$ represents the proportion of fully cooperative bots, while $1-\theta$ indicates the proportion of fully defective bots. Initially, bots and ordinary players are randomly distributed across network nodes. Unless otherwise specified, the strategic values for bots are fixed at $s_{BP}=0.9$, indicating that we are considering cooperative bots. The initial strategic values $s_{OP}^{initial}$ of ordinary players are randomly distributed between 0 and 1.

We also consider two distinct population types: well-mixed and network. In the well-mixed population, players interact with all other players with equal probability. This scenario is represented by a fully connected network, where each player is connected to every other, reflecting the interactive nature of well-mixed populations. Conversely, in the networked population, individuals interact only with their immediate neighbors. This scenario is represented by a regular two-dimensional (2D) lattice network with 8 nearest neighbors (top, bottom, left, right, top-left, top-right, bottom-left, and bottom-right), resulting in an average degree $\langle k \rangle$ of 8 for the network. The size of the network (population) is $N$. To ensure that all players in the network have the same number of interacting neighbors, we consider the network to have periodic boundaries.

\subsection{\label{sec:level2}Strategy update}
Since we aim to explore how bots can influence the cooperation levels of ordinary players even with the simplest settings, we assume that the bots are pre-programmed with specific strategies and never alter them. In contrast, ordinary players in real-world scenarios often adopt strategies through social learning by imitating the most successful players, though they occasionally make irrational decisions. Therefore, we assume that ordinary players adopt the pair-wise Fermi (PW-Fermi) as their strategy update rule. The specific details are as follows: player $i$ accumulates a payoff $\pi_i$ after playing the games with its closest $\langle k \rangle$ neighbors. Subsequently, it randomly selects player $j$ from its neighbors as its opponent. Player $j$ repeats the above process and obtains cumulative payoff $\pi_j$. Next, player $i$ will imitate the strategy of player $j$ with a probability $P$ as follows:
\begin{equation}
    P_{s_i\gets{s_j}}=\frac{1}{1+exp[{(\pi_i-\pi_j)\cdot{\kappa^{-1}}}]}
\end{equation}
where $\kappa^{-1}$ represents imitation strength, indicating how strictly the player decides whether to imitate their neighbors' strategies based on the difference in their payoffs~\cite{sigmund2010social,szabo1998evolutionary}. As $\kappa^{-1} \rightarrow +\infty$ (or when $\pi_i = \pi_j$), the decision to imitate or not is effectively determined by a coin toss, which we define as the `weak imitation' scenario. In this scenario, imitation occurs largely at random, although players with higher payoff tend to be imitated slightly more frequently. Conversely, when $\kappa^{-1} \rightarrow +\infty$, known as the `strong imitation' scenario, players consistently imitate those who are more benefit and avoid imitating those who have fewer payoffs. 
Unless otherwise specified, $\kappa^{-1}$ is fixed at 10 in this study, denoting the scenario of strong selection. In this case, the players' decisions are rational in the vast majority of situations.

\subsection{\label{sec:level2}Simulation settings}
For the well-mixed population, we set $N=500$; for the networked population, we set $N=10^{5}$. Both population structures conducted $10^4$ Monte Carlo simulations (MCS), with each player updating their strategy once on average per MCS. Our results are obtained by averaging the results from the last 2000 time steps, ensuring the
systems reached a stationary state after sufficiently long relaxation times. To enhance the reliability of our findings, we independently conducted 100 realizations for each case and obtained the averages.
It is noteworthy to mention that, given our focus on investigating the impact of bots on the cooperative behavior of ordinary players, the calculated probability of choosing cooperation ($F_C$) excludes the former and solely accounts for the latter.

\begin{figure}[!t]
    \centering
\includegraphics[width=0.96\linewidth]{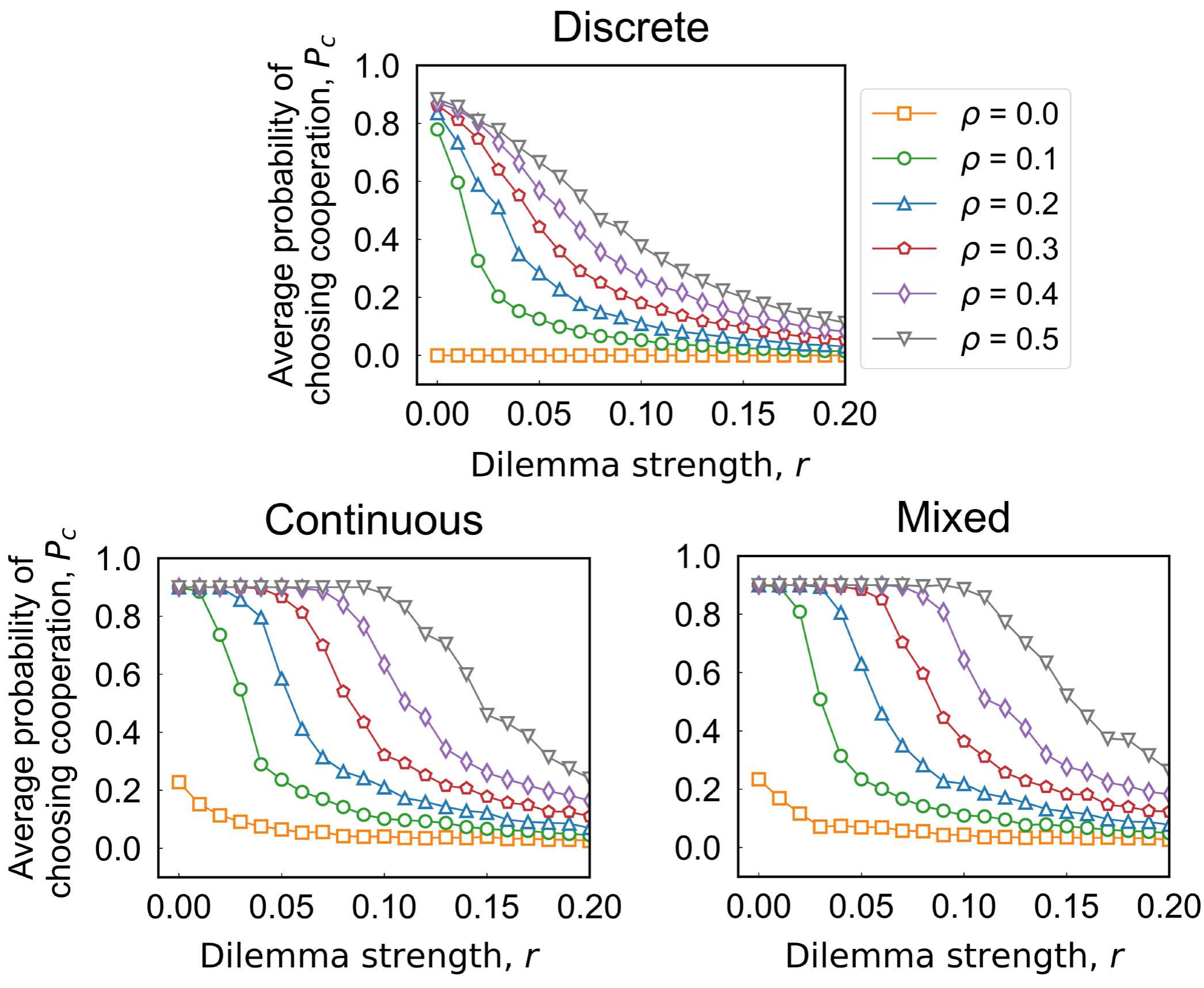}
      \caption{\textbf{In well-mixed populations, regardless of the strategic approaches adopted, introducing cooperative bots can enhance the cooperation levels among ordinary players.} Shown are the cooperation levels of ordinary players within well-mixed populations independent of the dilemma strength $r$ under the different proportions $\rho$ of bots. We examined three strategic scenarios: discrete (left panel), continuous (middle panel), and mixed (right panel), with $\rho$ varying from 0 to 0.5 and the $r$ ranging from 0 to 0.2. Imitation strength was fixed at $\kappa^{-1}=10$.}
      \label{fig1}
\end{figure}

\section{Results}
\subsection{\label{sec:level2}Well-mixed populations}

In well-mixed populations, defection is the strict Nash equilibrium in the prisoner's dilemma game under the discrete strategic approach, regardless of the dilemma strength $r$. Although continuous and mixed strategies do not alter the defective equilibrium, they introduce a degree of uncertainty in the evolution of cooperation in finite populations, eventually leading to the emergence of cooperation~\cite{si2024mixed}.

Compared to the discrete strategic framework, introducing cooperative bots among human players (hereafter referred to as ordinary players) within the continuous and mixed strategic approaches greatly enhances the cooperation-promotion effect by cooperative bots. As shown in Figure \ref{fig1}, although a large proportion of cooperative bots always facilitates cooperation levels among ordinary players within the discrete strategic approach, the cooperation level easily breaks down with increasing dilemma strength $r$. In contrast, within the continuous and mixed strategic approaches, cooperative bots can enforce the dominance of cooperation over a wide range of dilemma strength $r$ (i.e., $r \lesssim 0.1$ when the proportion of cooperative bots is 0.5).

\begin{figure}[!t]
    \centering
    \includegraphics[width=0.96\linewidth]{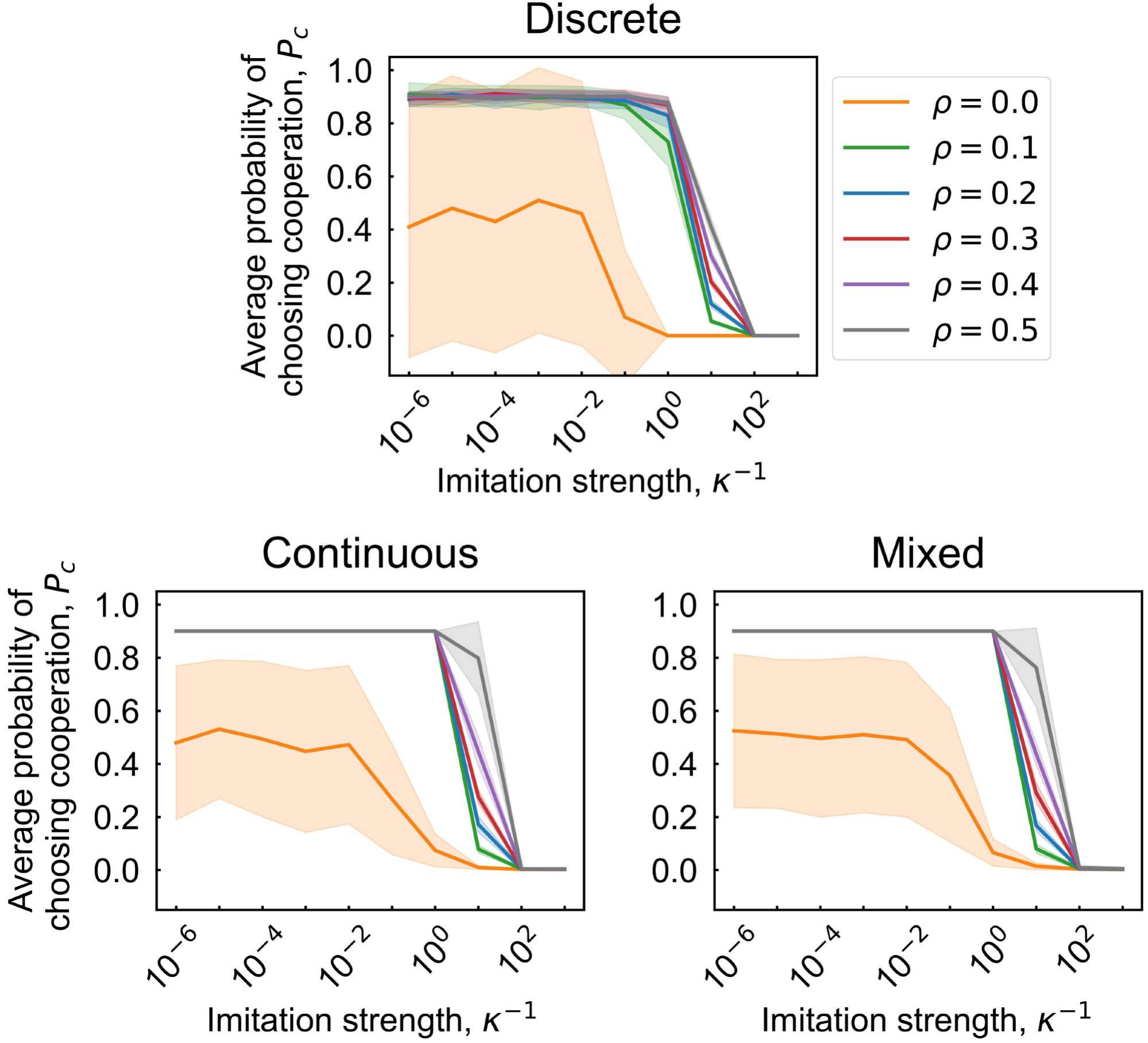}
      \caption{\textbf{The effect of cooperative bots on promoting cooperation in well-mixed populations is limited to scenarios with weak imitation strength across all three strategic approaches.} Shown are the cooperation levels of ordinary players within a well-mixed population independent of the imitation strength $\kappa^{-1}$ under the different proportions $\rho$ of bots. From left panel to right panel, we consider three strategic approaches: discrete, continuous, and mixed. The shaded areas indicate the variance between different simulation results. The dilemma strength was fixed at $r=0.1$.}
    \label{fig2}
\end{figure}

Further considering the influence of imitation strength on the effectiveness of cooperative bots on cooperation levels among ordinary players, we observe that the cooperation-promotion effect of cooperative bots is restricted when $\kappa^{-1} \lesssim 10^{2}$ across the three strategic approaches (Figure \ref{fig2}). Beyond this threshold, introducing cooperative bots does not alter the defective equilibrium. Additionally, we noted that introducing cooperative bots can mitigate the uncertainty of cooperation evolution, which typically occurs under weak imitation scenarios (i.e., the range of  $\kappa^{-1}$ where cooperation can survive in the absence of bots). Under weak imitation scenarios, players place less emphasis on their material gains, and evolutionary outcomes are influenced more by action frequencies than by material payoffs. Therefore, in the absence of bots, the final cooperation level exhibits a degree of unpredictability across the three strategic approaches. Although introducing cooperative bots cannot alter the defective equilibrium, it increases the frequencies of cooperation in the hybrid populations, giving ordinary players more opportunities to encounter cooperators. Consequently, introducing cooperative bots not only mitigates the uncertainty of cooperation evolution but also enhances cooperation among ordinary players across these strategic approaches.

Overall, the results confirm the positive role of cooperative bots in enhancing cooperation among ordinary players under weak imitation scenarios within the discrete strategic approach~\cite{masuda2012evolution,guo2024engineering,guo2023facilitating}. Additionally, they reveal an enhanced cooperation-promotion effect by cooperative bots under continuous and mixed strategic approaches. Given that existing studies have shown the destructive effect of cooperative bots on cooperation in structured populations within the discrete strategic framework~\cite{matsuzawa2016spatial}, we next turn our attention to structured populations to investigate the role of cooperative bots on cooperation among ordinary players within the continuous and mixed strategic frameworks.

\subsection{\label{sec:level2}Structured populations}

\begin{figure}[!t]
    \centering
    \includegraphics[width=0.96\linewidth]{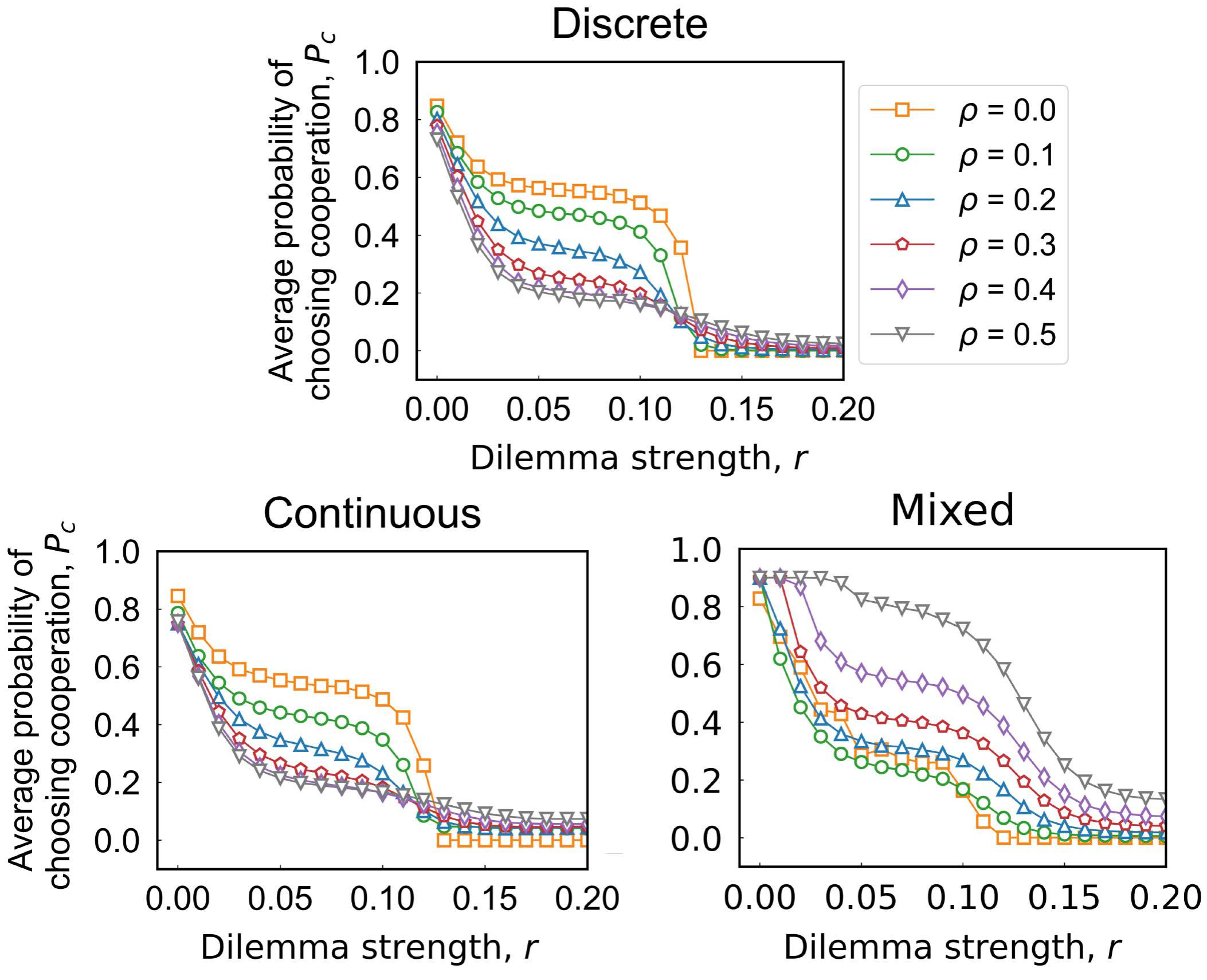}
      \caption{\textbf{Unlike a well-mixed population, bots under discrete and continuous strategic approaches in a network population suppress the level of cooperation among ordinary players, but the mixed strategic approach still promotes it.} Shown are the cooperation levels of ordinary players within a networked population independent of the dilemma strength $r$ under the different proportions $\rho$ of bots. We examined three strategic scenarios: discrete (left panel), continuous (middle panel), and mixed (right panel), with $\rho$ varying from 0 to 0.5 and the $r$ ranging from 0 to 0.2. Imitation strength was fixed at $\kappa^{-1}=10$.}
      \label{fig3}
\end{figure}

Unlike the cooperation-promoting effect observed with cooperative bots in well-mixed populations, introducing cooperative bots into the Prisoner's Dilemma game on a regular lattice suppresses cooperation among ordinary players within the framework of discrete and continuous strategic approaches. This suppression occurs even when network reciprocity alone, which allows cooperators to form compact clusters that support each other, favors cooperation (i.e., $r<0.13$, Figure~\ref{fig3}, left and middle panel). Although there is a slight increase in cooperation levels among ordinary players when introducing cooperative bots in scenarios where network reciprocity alone cannot favor cooperation (i.e., $r>0.13$), this increment is minimal, with the maximum cooperation level remaining below 10\% even when the fraction of cooperative bots reaches 0.5. In contrast to the detrimental effects seen under discrete and continuous approaches, the mixed strategic approach allows cooperative bots to enhance cooperation levels among ordinary players. This promotional effect strengthens as the proportion of cooperative bots increases, regardless of the dilemma strength (refer to Figure~\ref{fig3}, right panel).

\begin{figure}[!t]
    \centering
    \includegraphics[width=0.96\linewidth]{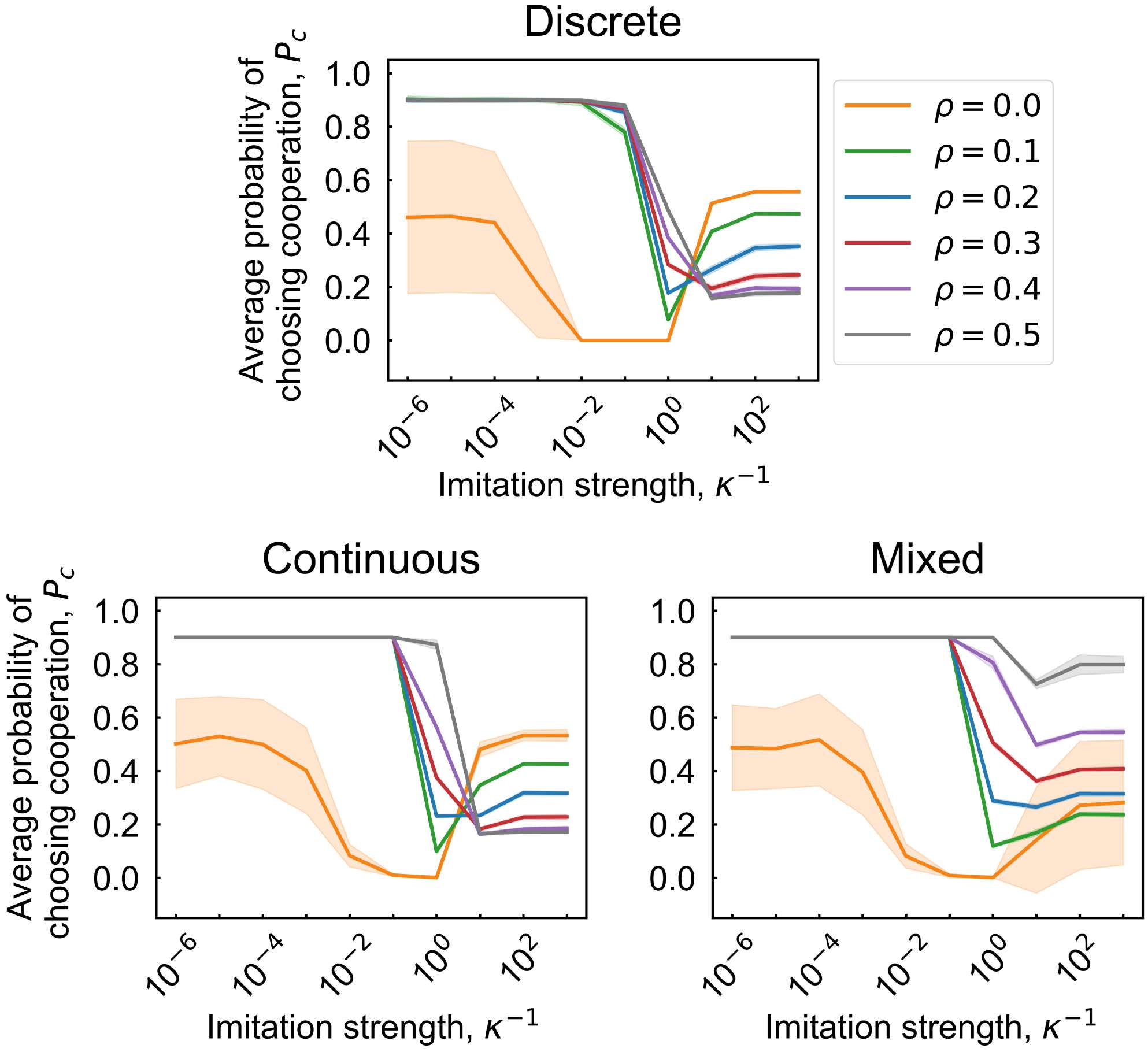}
      \caption{\textbf{In a regular lattice, under discrete and continuous strategic approaches, introducing bots affects cooperation levels among ordinary players differently depending on imitation strengths. However, under the mixed strategic approach, introducing bots almost always enhances cooperation among ordinary players regardless of imitation strength.} Shown are the average cooperation fractions of ordinary players, obtained from averaging over 100 simulations, as a function of the imitation strength within a networked population. From left panel to right panel, we consider three strategic approaches: discrete, continuous, and mixed. The solid lines of different colors represent the different proportions of bots. The shaded areas indicate the variance between different simulation results. All other parameters are consistent with Figure~\ref{fig2}.}
    \label{fig4}
\end{figure}

Figure~\ref{fig4} further examines the role of imitation strength on the effectiveness of cooperative bots in enhancing cooperation levels among ordinary players across the three strategic approaches within a square lattice. It can be seen that the distinct impact of cooperative bots on cooperation among ordinary players across these strategic approaches in structured populations is restricted to strong imitation scenarios (i.e., $\kappa^{-1} \geq 10^0$). Consistent with the effects observed in well-mixed populations, the cooperation-promotion effect of cooperative bots remains valid in weak imitation scenarios across these strategic approaches, and they also mitigate the uncertainty of cooperation evolution.

\begin{figure*}[ht]
    \centering
\includegraphics[width=0.91\linewidth]{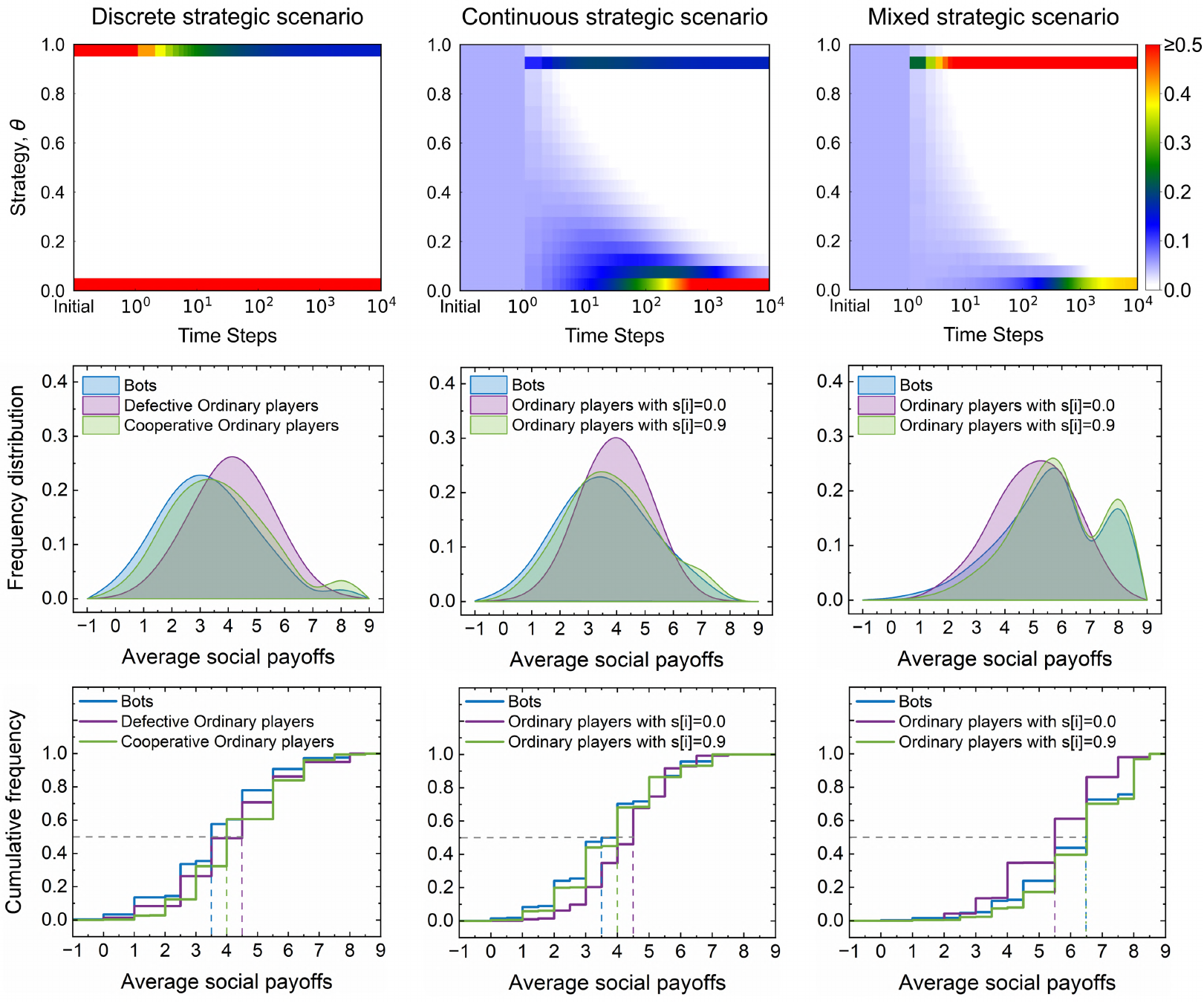}
      \caption{\textbf{Mixed strategic approach allows ordinary players with high probabilities of choosing cooperation (high values of strategies) to opt for defection, thereby achieving higher payoffs.} Shown in the top row are the time evolution diagrams of the frequency for ordinary players with different strategy values. The middle and bottom rows, respectively, show the frequency distribution and cumulative frequency of the average social payoffs for bots, ordinary players with a strategy value of 0, and ordinary players with a strategy value of 0.9 during the total time steps. Moving from left to right, we employ discrete, continuous, and mixed strategic approaches. The parameter settings are as follows: the proportion of bots in the hybrid group was fixed at $\rho=0.5$; the dilemma strength was fixed at $r=0.1$; and the imitation strength was fixed at $\kappa^{-1}=10$.}
      \label{fig5}
\end{figure*}

Under strong imitation scenarios, why do cooperative bots enhance cooperation among ordinary players within the mixed strategic approach but not under discrete and continuous strategic approaches? Given that mixed and continuous strategies are similar—both allowing the probability of choosing cooperation—the question is crucial. In the continuous strategy, players receive expected payoffs based on their cooperation probability. In contrast, mixed strategies require players to choose between cooperation and defection based on their probability of cooperation, resulting in immediate payoffs. We hypothesize that the differing impacts of cooperative bots on ordinary players under these strategies stem from differences in payoff calculation methods.
To test this hypothesis, we illustrated the findings in Figure \ref{fig5}. Under the discrete (left panel) and continuous (middle panel) strategic approaches, cooperative bots have the lowest average social payoffs, followed by ordinary players who choose to cooperate, while ordinary players who choose to defect have the highest average social payoffs (the payoff distribution of these players generally follows normal distributions). This results in ordinary players gradually decreasing their cooperation levels to maximize their personal gains (top row, left two panels of Figure \ref{fig5}). However, under the mixed (right panel) strategic approach, although cooperative bots mostly choose to cooperate and thus receive the lowest payoffs, they still have a small probability of choosing defection (e.g., 10\% chance of defecting). This stochastic feature inherent in the mixed strategic approach allows cooperative bots to occasionally achieve payoffs that exceed those of ordinary players who defect. Consequently, the payoff distributions of cooperative bots and cooperative ordinary players follow a mixture of binormal distributions (middle row, right panel of Figure \ref{fig5}). This feature leads to a bifurcation for ordinary players, who evolve either towards the strategy of the bots or towards pure defection (refer to Figure \ref{figa1}), ultimately leading to the polarization of ordinary players in their strategy over time (top right panel of Figure \ref{fig5}).

\begin{figure}[!t]
    \centering
    \includegraphics[width=0.96\linewidth]{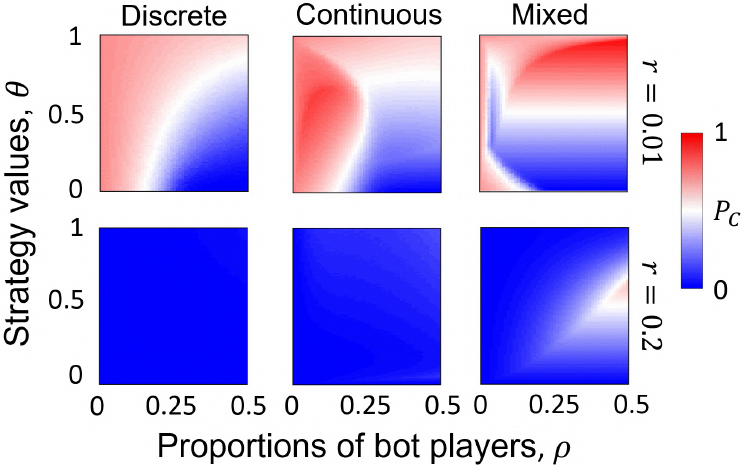}
      \caption{\textbf{In a regular lattice, as opposed to discrete and continuous strategic approaches, the mixed strategic approach enables achieving a high level of cooperation, even under high dilemma strengths.} Shown are the variations in cooperation levels among ordinary players as a function of bots' proportion and strategy values in a regular lattice. The top row to the bottom row represents dilemma strengths $r = 0.01, 0.2$. The columns from left to right correspond to the strategic approaches of discrete, continuous, and mixed. The imitation strength was fixed at $\kappa^{-1}=10$.}
      \label{fig6}
\end{figure}

In the scenarios previously discussed, we only considered cooperative bots with a high probability of choosing to cooperate, leaving the question of how varying this probability affects their influence on cooperation levels among ordinary players undetermined. To address this, we examined the scenario under strong imitation strength ($\kappa^{-1}=10$) and illustrated the results in Figure \ref{fig6}. The top row in Figure \ref{fig6} considers weak dilemma strength ($r=0.01$), where network reciprocity alone can support the emergence of cooperation among ordinary players. We observe that, compared to defective bots ($\theta < 0.5$), cooperative bots generally promote cooperation across all three strategic approaches. However, there are optimal levels of $\theta$ where the cooperation level among ordinary players is highest, particularly under continuous and mixed strategic approaches. This optimal value of $\theta$ can also be observed under mixed strategic approaches when network reciprocity alone cannot support the emergence of cooperation among ordinary players (bottom row of Figure \ref{fig6}). These results indicate that achieving optimal cooperation levels among ordinary players does not necessarily require perfectly cooperative bots; rather, a degree of stochastic behavior (i.e., behavior noise) plays a decisive role in enforcing cooperation. This finding aligns with our earlier results shown in Figure \ref{fig5}, explaining why cooperative bots can still enforce cooperation compared to the other two approaches.

\section{Conclusion and discussion}
To discuss, we extended the investigation of the effectiveness of cooperative bots to the frameworks of continuous and mixed strategic approaches. While previous studies have recognized that the effectiveness of cooperative bots on cooperation among ordinary players within discrete strategic frameworks is limited to weak imitation scenarios~\cite{guo2023facilitating,guo2024engineering,cardillo2020critical,masuda2012evolution}, their role within continuous and mixed strategic approaches has been largely unexplored. Our findings reveal the distinct impact of cooperative bots on cooperation levels among ordinary players across these strategic approaches in structured populations under strong imitation scenarios. Specifically, cooperative bots devastate network reciprocity and lead to the breakdown of cooperation within discrete and continuous strategic approaches but still promote cooperation in mixed strategic approaches. Taken together, these results highlight the nuanced effects of cooperative bots on human cooperation across different strategic frameworks, enriching our understanding of how cooperative bots influence cooperation within evolutionary game theory. On the other hand, these results also underscore the need for careful deployment of cooperative bots to stimulate human cooperation, as their effectiveness is highly sensitive to how humans update their actions and their chosen strategic approach.

Discrete, continuous, and mixed strategic approaches are fundamental aspects of evolutionary game theory. Although these strategic approaches do not alter the defective equilibrium and only differ in how players make their actions and calculate their payoffs (using deterministic or stochastic patterns under given probabilities), they exhibit distinct degrees of uncertainty in the direction of cooperation evolution, leading to different cooperation levels within finite human-human populations~\cite{si2024mixed}. In human-machine scenarios, introducing cooperative bots mitigates the uncertainty in the direction of cooperation evolution across all strategic approaches. The inherent stochastic nature of the mixed strategic approach allows cooperative bots to occasionally achieve high payoffs through defecting, which enhances network reciprocity and promotes cooperation. This contrasts with the discrete and continuous approaches, where cooperative bots rarely achieve high payoffs and instead create a breeding ground for defectors, leading to a decline in cooperation. Our results on human-machine interactions, combined with findings from human-human interactions, provide a deeper understanding of the role of these three strategic approaches in cooperation problems. Additionally, the distinct role of cooperative bots in promoting or hindering cooperation across these strategic approaches serves as a valuable starting point for extending the model to more complex scenarios, such as repeated interactions~\cite{wang2017onymity,wang2018exploiting} or one-shot games with varying levels of information~\cite{nowak2005evolution,gintis2001costly}.

The concept of fixed-behavior AI is similar to that of zealots, who consistently engage in unconditional cooperation and never change their actions~\cite{shen2023committed}. Fixed-behavior AI extends this concept by encompassing individuals committed to a single strategy, guiding either their cooperative or defective actions, and varies in its applicability~\cite{sharma2023small}. If a significant number of zealots are required to foster cooperation, their impact might become negligible. In the era of AI, people increasingly interact with AI entities controlled by various algorithms in the digital world. In our one-shot and anonymous game framework, which excludes cooperation-supporting mechanisms, the distinction between these concepts may be blurred. However, existing studies have revealed the phenomenon of machine penalty, suggesting that awareness of opponents' identities could influence cooperation. Therefore, further research could abandon the anonymity assumption and incorporate people's emotions~\cite{erol2019toward,de2019cooperation}, social preferences~\cite{dafoe2021cooperative,mckee2020social}, psychological aspects~\cite{hoc2001towards}. Moreover, more realistic scenario settings employing large language models with well-crafted prompts can better depict human decision-making in complex, realistic scenarios~\cite{ren2024emergence}. This approach will more thoroughly investigate how cooperative bots affect human cooperation within the framework of evolutionary game theory.

\begin{acknowledgments}
We acknowledge support from (i) JSPS KAKENHI (Grant no. JP 23H03499) to C.\,S., (ii) China Scholarship Council (Grant no.~202308530309) and Yunnan Provincial Department of Education Science Research Fund Project (Grant no. 2024Y503) to Z.H., and (iii) the grant-in-Aid for Scientific Research from JSPS, Japan, KAKENHI (Grant No. JP 20H02314 and JP 23H03499) awarded to J.\,T.
\end{acknowledgments}

\section*{Author contribution}
C. S. and J. T. conceived research. Z. S. and Z. H. performed simulations. All coauthors discussed the results and wrote the manuscript.

\section*{Conflict of interest}
Authors declare no conflict of interest.

\section*{Data Availability Statement}

The code used in the study to produce all results is freely available at \url{https://osf.io/tp52h/}.

\section*{Appendix}
In Figure \ref{figa1}, we provide an example to illustrate how bots with specific strategies adopting two different actions can drive ordinary players to evolve in different directions under mixed strategic scenario. Consider a focal ordinary player with a strategy value of $S_{FOP}=0.4$, surrounded by bots with $S_{BP}=\theta$ and ordinary players with $S_{OP}=0$. The focal ordinary player observes the following situation:
The left neighbor is a bot with $S_{BP}=\theta$ who chooses the defection action during the game. The right neighbor is an ordinary player with $S_{OP}=0$ who also chooses the defection action during the game, and among her eight closest neighbors, four are bots. Due to the stochastic nature of the bots' choice between cooperation and defection, subtle differences in the payoffs of the focal ordinary player’s neighbors arise in different episodes:

(a):
Out of the four bot neighbors of the right neighbor, two choose to cooperate, and two choose to defect.
As a result, the right neighbor obtains a cumulative payoff of $2(1+r)$.
The left neighbor obtains a cumulative payoff of $3(1+r)$.
Since $2(1+r) < 3(1+r)$, the focal ordinary player imitates the left neighbor's strategy, adopting $S_{FOP}=S_{BP}=\theta$.

(b):
All four bot neighbors of the right neighbor choose to cooperate.
Therefore, the right neighbor obtains a cumulative payoff of $4(1+r)$.
The left neighbor obtains a cumulative payoff of $3(1+r)$.
Since $4(1+r) > 3(1+r)$, the focal ordinary player imitates the right neighbor's strategy, adopting $S_{FOP}=S_{OP}=0$.

\label{sec:Appendix}
\setcounter{equation}{0}
\renewcommand\theequation{A\arabic{equation}}
\renewcommand\thefigure{A\arabic{figure}}
\setcounter{figure}{0}  
\begin{figure}[!t]
    \centering
\includegraphics[width=0.91\linewidth]{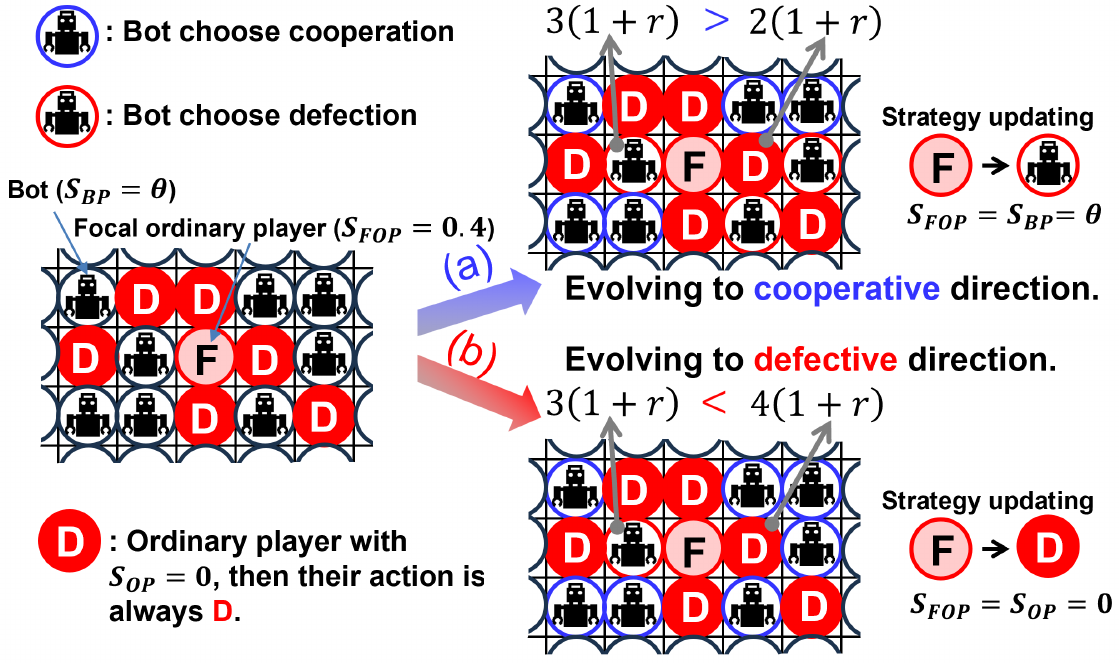}
      \caption{\textbf{Mixed strategic approach allows bots to choose different actions (cooperation or defection) in various episodes and obtain different payoffs, thereby influencing the evolutionary direction of ordinary players.} Shown is the schematic diagram of bots with specific strategies adopting two different actions that can drive ordinary players to evolve in different directions under mixed strategic scenario. }
    \label{figa1}
\end{figure}

\clearpage

\bibliographystyle{elsarticle-num}
\bibliography{biblio}

\end{document}